\documentclass[letterpaper, 10 pt, conference]{ieeeconf}  

\IEEEoverridecommandlockouts                              

\usepackage{epsfig} 
\usepackage{mathptmx} 
\usepackage{times} 
\usepackage{amsmath} 
\usepackage{amssymb}  

\usepackage[utf8]{inputenc} 
\usepackage[T1]{fontenc}    
\usepackage{url}            
\usepackage{booktabs}       
\usepackage{amsfonts}       
\usepackage{nicefrac}       
\usepackage{microtype}      
\usepackage{xcolor}         

\usepackage{graphicx}
\usepackage{amsthm}
\usepackage{ragged2e} 
\usepackage{stackengine}
\usepackage{bbm}

\newcommand{\bR}{\mathbb{R}}
\newcommand{\bP}{\mathbb{P}}
\newcommand{\bE}{\mathbb{E}}
\usepackage{forest}

\usepackage{mathtools}

\newcommand{\supp}{\mathrm{supp}}

\newcommand{\calD}{{\mathcal{D}}}
\newcommand{\calR}{{\mathcal{R}}}
\newcommand{\calF}{{\mathcal{F}}}
\newcommand{\range}{\mathrm{range}}

\DeclarePairedDelimiter\abs{\lvert}{\rvert}%
\DeclarePairedDelimiter\norm{\lVert}{\rVert}%

\newtheorem{theorem}{Theorem}

\newtheorem{definition}{Definition}
\newtheorem{proposition}{Proposition}
\newtheorem{assumption}{Assumption}
\newtheorem{lemma}{Lemma}

\newtheorem{remark}{Remark}

\title{\LARGE \bf
Learning of Dynamical Systems under Adversarial Attacks - Null Space Property Perspective}

\author{Han Feng, Baturalp Yalcin, and Javad Lavaei
\thanks{This work was supported by grants from ARO, AFOSR, ONR, and NSF. The authors are with the University of California, Berkeley. E-mail:}
\thanks{ 
        {\tt\small \{han\_feng, byalcin, lavaei\}@berkeley.edu}}%
}

\begin{document}

\maketitle
\thispagestyle{empty}
\pagestyle{empty}

\begin{abstract}

We study the identification of a linear time-invariant dynamical system affected by large-and-sparse disturbances modeling adversarial attacks or faults. Under the assumption that the states are measurable, we develop necessary and sufficient conditions for the recovery of the system matrices by solving a constrained lasso-type optimization problem. In addition, we provide an upper bound on the estimation error whenever the disturbance sequence is a combination of small noise values and large adversarial values. Our results depend on the null space property that has been widely used in the lasso literature, and we investigate under what conditions this property holds for linear time-invariant dynamical systems. Lastly, we further study the conditions for a specific probabilistic model and support the results with numerical experiments.

\end{abstract}

\section{INTRODUCTION}

The identification of linear time-invariant (LTI) systems is a classic problem in control theory that has been studied extensively. Despite the long history of this problem and its application in a wide range of real-world systems, the non-asymptotic analysis of the system identification problem has gained popularity in recent years, which targets the sample complexity of the problem \cite{dean2018regret,fattahi2020efficient}. With the growing popularity of safety-critical applications, such as autonomous driving and unmanned aerial vehicles, the design of a system identification framework that is robust against adversarial attacks is crucial \cite{fisacGeneralSafetyFramework2019}.

In this paper, we consider LTI systems for which the states are under a sequence of unknown disturbances, some of which take small values modeling noise and the remaining ones take strategic values due to adversarial attacks or faults in the system. We study the lasso-based optimization problem recently proposed in \cite{hanprev}. It can recover the exact system dynamics uniquely whenever adversarial attacks occur intermittently with enough time separation. In \cite{hanprev}, some adversarial attacks that do not influence the estimation are studied, whereas this paper improves those results by providing a necessary and sufficient condition for recovery as well as  non-asymptotic bounds on the error. Our approach is based on defining a null space property that is analogous to the null space property condition for the lasso problem \cite{wainwright_2019}, which is required to guarantee the exact recovery.

The robustness analysis of estimators has a long history, dating back to the seminal paper \cite{tukeyFutureDataAnalysis1962}. It is known that a small disturbance on the estimation problem, such as the perturbation of a data point, could lead to significant changes in the outcome of the estimator. This has led to a major effort on the robustification estimators
via regularizers. The works  \cite{xuRobustnessRegularizationSupport2009} and \cite{bertsimasCharacterizationEquivalenceRobustification2018} have found a strong relationship between the robust estimation and regularization in regression problems by showing the equivalence of these two problems. 

The recent papers \cite{bakoClassOptimizationBasedRobust2017} and \cite{bakoAnalysisNonsmoothOptimization2016} on robust estimation of linear measurement models have considered a framework with two types of noise: small measurement noise and large intermittent noise. They have developed necessary and sufficient conditions for the exact recovery when a column-wise summable norm is used to minimize the error. We focus on this type of norm in this paper, which will be defined as the sum of $\ell_2$ norms of the columns of a matrix. Nevertheless, our analysis is for the more challenging problem of system identification where the parameters are correlated over time.
 Our results indirectly provide a guideline on how to design an effective input sequence to learn system dynamics faster.

The recent papers \cite{fattahiDataDrivenSparseSystem2018} and \cite{fattahi2018sample} on system identification have studied the problem of learning a sparse and structured state-space model, and provided bounds on the required sample size, i.e., sample complexity bounds. However, none of the aforementioned works are applicable to adversarial attacks since their noise/disturbance model is Gaussian. The more recent work \cite{molybogConicOptimizationRobust2018} has utilized a conic relaxation, which significantly increases the problem dimension and is not directly applicable to dynamical systems. It estimates how many erroneous measurements or adversarial attacks can be handled by the estimator without causing a nonzero estimation error. There are some other works in the literature that provide non-asymptotic error bounds for the linear system identification problem when the ordinary least-squares estimation method and Kalmon-Ho algorithm are used \cite{oymakNonasymptoticIdentificationLTI2019, sarkarOptimalFiniteTime2019}. However, these methods are not particularly efficient for robust estimation whenever the data is corrupted in an adversarial way. Membership estimators are also utilized to show a consistent estimation of linear systems \cite{hespanholStatisticalConsistencySetMembership2020a}. Unfortunately, they do not provide non-asymptotic bounds. The work \cite{showkatbakhshSystemIdentificationPresence2016a} has studied the scenario where the attack is executed on the outputs rather than the states. Unlike the attack on the outputs, the effect of the attack on the states propagates over time. Lastly, some other related works on robust estimation are resilient state estimation~\cite{fawziSecureEstimationControl2014,chongObservabilityLinearSystems2015} and Byzantine fault tolerance~\cite{suFiniteTimeGuaranteesByzantineResilient2020,guptaFaultToleranceDistributedOptimization2020}.

One could place the system identification problem into the broader context of robust regression to gain some valuable insight on robust estimation. It is well-known that least-squares methods are not robust to outliers. The work \cite{sheOutlierDetectionUsing2011} has studied the identification of outliers in linear regression. It is shown that a non-convex loss function outperforms the $\ell_1$ regularization of the least-squares function. Nevertheless, it is not always justifiable to solve large-scale non-convex problems instead of convex ones unless the landscape of the non-convex optimization problem can be shown to be benign (e.g., it does not have a spurious solution). Nonetheless, this is problem specific and not understood thoroughly \cite{joszTheoryAbsenceSpurious2018,molybog2020role}. Another mainly used estimator for sparse estimation is the hard thresholding estimator. The work \cite{bhatiaConsistentRobustRegression2017} has proposed an iterative scheme based on this estimator and analyzed it on regression with sparse disturbances. There have been several other works on robust estimation and training \cite{diakonikolasSeverRobustMetaAlgorithm2019,prasadRobustEstimationRobust2020a,steinhardtCertifiedDefensesData2017}. The major difference between those works and the system identification is that the states or the training data are not independent over time. Hence, they cannot be re-ordered, which makes it challenging to exploit the existing results in robust statistics. A possible solution to this is resetting the system and using the last available data point from each trajectory. However, this is not a feasible approach for common real-life applications due to its complexity. Also, it is often desired to identify the system in an online fashion to benefit from the available data, but it is not well understood how this can be achieved for robust estimation. 

In Section \ref{sec:notations}, we introduce the main notations used in the paper. Section~\ref{sec:problem_formulation} considers a particular type of $l_1$ minimization problem and formulates the problem. In Section~\ref{sec:noiseless}, we derive sufficient conditions for exact recovery in finite time when we have exact measurements of the states that are influenced by the adversarial attacks. The noisy case is studied in Section~\ref{sec:noisy}, where we provide an error bound on the estimation error based on the noise intensity. The conditions are based on the null space property (NSP), which is hard to verify directly. We derive sufficient conditions for NSP in Section~\ref{sec:nsp} and then show in Section~\ref{sec:prob} that NSP holds for a particular attack model where the input is Gaussian and the adversary injects disturbances intermittently with a fixed policy based on the states and input measurements. The proofs are provided in the appendix. 

\section{NOTATIONS} \label{sec:notations}

For a given matrix $Z$, the $i$-th largest singular value of $Z$ is denoted by $\sigma_i(Z)$, and the minimal and maximum singular values of $Z$ are shown by $\sigma_{\min}(Z)$ and $\sigma_{\max}(Z)$, respectively. For a matrix $Z$, $\| Z \|_F$ denotes its Frobenius norm and for a vector $z$, $\| z \|_2$ denotes its $\ell_2$ norm. $Z \succ 0$ and $Z \succeq 0$ denote a square symmetric matrix $Z$ that is positive definite and positive semidefinite, respectively. The function $\text{tr}(\cdot)$ stands for the trace of a square matrix.  The $n \times n$ identity matrix is denoted as $\mathbf{I}_n$. The Minkowski sum of two sets $\mathcal{E}$ and $\mathcal{F}$ is denoted by $\mathcal{E}\oplus\mathcal{F} = \{e + f :  e\in \mathcal{E}, f\in \mathcal{F}\}$. The sum with the inverse of the set is denoted by $\mathcal{E}\ominus\mathcal{F} = \{e - f: e\in \mathcal{E}, f\in \mathcal{F}\}$. For two vectors $v$ and $w$, $\langle v, w \rangle$ is the inner product between those vectors in their respective vector space. $\bP(\cdot)$ and $\bE[\cdot]$ denote the probability of an event and the expectation of a random variable. A Gaussian random variable $X$ with mean $\mu$ and covariance matrix $\Sigma$ is written as $X \sim N(\mu, \Sigma)$. $| S |$ shows the cardinality of a given set $S$.

\section{PROBLEM FORMULATION}
\label{sec:problem_formulation}

Consider an LTI dynamical system over the time horizon $[0, T]$:
\begin{align*}
  x_{t+1} = \bar A x_t + \bar Bu_t + \bar d_t,  \quad t=0, 1, \ldots, T-1,
\end{align*}
where $\bar A \in \bR^{n\times n}$ and $ \bar B\in \bR^{n\times m}$ are unknown matrices in the state-space model to be estimated and $\bar d_t$'s are unknown disturbances. Throughout the paper, the bar over each parameter of interest (such as $\bar A$) indicates the unknown ground truth. The goal is to find the matrices $\bar A$ and $\bar B$ from the state measurements $x_0,...,x_T\in\mathbb R^n$ and input data $u_0,...,u_{T-1}\in\mathbb R^m$.
The disturbances $\bar d_0,...,\bar d_{T-1}$ model both noise and anomalies in the system, such as attacks or actuator's faults.
Without any assumptions on the disturbance, the identification problem is not well-defined due to the impossibility of separating $\bar Ax_t + \bar Bu_t$ from the disturbance $\bar d_t$. For instance, if $\bar d_t = A' x_t + B' u_t$ for some matrices $A'$ and $B'$, then the system evolves as if the system matrices are $(\bar A + A', \bar B + B')$ and the disturbance is zero, which makes the identification problem have non-unique solutions. We will make certain sparsity assumptions on the disturbance in the noiseless case, and generalize the result to the noisy case.

To formulate the problem, we introduce the matrix notation $X = [x_0, \ldots, x_{T-1}]$, $U  = [u_0, \ldots, u_{T-1}]$, and $D=[d_0, \ldots, d_{T-1}]$. The last state $x_T$ appears in our optimization problem, but it is not a column in the matrix notation. The attack $D$ is assumed to be restricted to a set $\calD \subseteq \bR^{n \times T}$. The set $\calD$ captures the user's belief of possible times of attack and their values. 

Define the sum of norm error $\norm{D}_{2, col}:= \sum_i \norm{d_i}_2$, where the index is over the columns of $D$.
The (column-wise) support of $D$ is defined as $\supp(D) = \{i \in \{0, \ldots, T-1\}: d_i \neq 0\}$. For each subset of indices $I\subseteq \{0, 1, \ldots, T-1\}$, the complement of $I$ is defined as $I^c = \{i \in \{0, \ldots, T-1\}: i\notin I\}$. For a matrix $Z\in \bR^{n\times T}$, the \emph{projection} $\Pi_I Z$ is a matrix whose columns are zero except for those in $I$, i.e.,
\[
  (\Pi_I Z)_i = \begin{cases}
    z_i, & \text{ if } i \in I \\
    0,   & \text{ otherwise}
  \end{cases},
\]
where $z_i$ denotes the $i$-th column of $Z$. Define $Z_I$ as a submatrix of $Z$ of size $n \times \abs{I}$, that includes only those columns of $Z$ in the index set $I$.
We use the shorthand notations $Z_{\neq i}$ and $Z_{\not \in I}$ to denote $Z_{\{0, \ldots, T-1\} \setminus \{i\}}$ and $Z_{\{0, \ldots, T-1\} \setminus I}$, respectively. The range of $Z$ is defined as $\mathcal{R}({Z}) = \{\sum_i \lambda_i z_i: \lambda_i \in \bR\}$.

To recover the system matrices $A$ and $B$, we analyze the following convex optimization problem:
    \begin{align}
  \min_{ \substack{A \in \mathbb{R}^{n \times n}, B \in \mathbb{R}^{n \times m}, \\ D\in \calD} }\;\, & \sum_{i=0}^{T-1} \norm{d_i}_2  \label{eq:hard-input-lasso}                      \\
  s.t. \quad                   & x_{i+1} = A x_i + B u_i + d_i, \quad {i=0, \ldots, T-1}, \notag
\end{align}
where the states $x_i, i\in \{0, \ldots, T\}$, are generated according to
\begin{align}
  x_{i+1} = \bar A x_i + \bar B u_i + \bar d_i, \quad {i=0, \ldots, T-1}. \label{eq:generate-input}
\end{align}
The control inputs $u_i, i\in \{0, \ldots, T-1\}$, may be designed but are fixed in the optimization problem \eqref{eq:hard-input-lasso}. This problem differs from the classical $l_1$ minimization (basis pursuit) problem that aims to find a ground truth vector $\bar z$ via 
\begin{align*}
  \min_z \quad & \norm{z}_1            \\
  s.t. \quad   & \Phi \bar z = \Phi z,
\end{align*}
for a given matrix $\Phi$ since
\begin{itemize}
  \item We apply the $l_1$ norm at the group level to the disturbances $d_1, \ldots, d_{T-1}$, because we only assume sparsity in the occurrence of the disturbance but not its value.
  \item The disturbance matrix $D$ is restricted to a set $\calD$.
  \item We do not attempt to minimize the $l_1$ norm of all the unknown parameters. In particular, the system matrices $A$ and $B$ are not assumed to be sparse.
  \item The states $x_0, \dots, x_{T-1}$ appear in the constraints and depend on the input $u_i$ and the disturbance $d_i$.
  \item Because the states are correlated, we cannot independently rescale them, as is commonly done in the analysis of $l_1$ optimization problems.
\end{itemize}

\section{THE NOISELESS CASE} 
\label{sec:noiseless}
This section studies the noiseless case, where each disturbance $\bar d_i$ is either zero or designed by an attacker to disturb the operation of the system. We aim to understand how to design the input of the system so that the identification of the excited system in the presence of adversarial disturbances is possible. We use $S = \supp(\bar D)$ to denote the time stamps of actual attacks. The set of possible disturbances $\mathcal D$ is assumed to be closed under the projection onto $S$.

\begin{assumption}
  The set of disturbances $\calD$ is convex and contains $0$ in its interior. Furthermore, $\Pi_S(D) \in \calD$ for all $D\in \calD$.
\end{assumption}

A key step in the study of problem \eqref{eq:hard-input-lasso} is the Null Space Property \cite{foucartMathematicalIntroductionCompressive2013}, which is formalized below.
\begin{definition}
  Let $c>0$, $S\subseteq \{0, \ldots, T-1\}$, and $\calR$ be a subset of $\bR^{n\times T}$.
  The matrix $\begin{bmatrix} X^T & U^T\end{bmatrix}^T \in \bR^{(n+m) \times T}$ is said to satisfy the Null Space Property (NSP) with the constant $c$, index set $S$, and range set $\calR$ ($(c, S, \calR)$-NSP) if, for all matrices $A \in \bR^{n \times n}$ and $ B \in \bR^{n \times m}$ such that 
  $- AX - BU \in \calR$ and $(A, B)$ is not zero, it holds that
  \begin{align}
    \left \| [A, B]    \begin{bmatrix}
        X_{S} \\ U_{S}
      \end{bmatrix}   \right\|_{2, col}  < c \left \|[A, B]     \begin{bmatrix}
        X_{S^c} \\ U_{S^c}
      \end{bmatrix}   \right\|_{2, col}. \label{eq:csnsp}
  \end{align}
\end{definition}
When the set $S$ or $\calR$ is clear from the context, we omit them and use $c$-NSP or $(c,S)$-NSP to highlight the parameters of interest. NSP has been used in \cite{candesStableSignalRecovery2006} to prove an exact recovery result and been further studied in \cite{cohenCompressedSensingBest2009}.
The following theorem formalizes a standard result that roughly states that 1-NSP is sufficient for the exact recovery of all sparse disturbances.
\begin{theorem}\label{thm:nsp-unique} The following statements are equivalent: 

\begin{itemize}
    \item[a)] $\begin{bmatrix} X\\ U\end{bmatrix}$ satisfies the $(1, S, \calD\ominus\calD)$-NSP where $S = \supp(\bar D)$.
    \item[b)] $(\bar A, \bar B, \bar D)$ is the unique solution to problem \eqref{eq:hard-input-lasso}.
\end{itemize}

\end{theorem}

The paper \cite{bakoClassOptimizationBasedRobust2017} has shown that 1-NSP is necessary for the exact recovery of all instances of a certain class of robust regression problems. However, because $x_0, \dots, x_{T-1}$ appear on both sides of the constraint, the system identification problem under study is structured and is only a subset of all instances of the regression problems.

\begin{remark}
  The case without control input is a special case of problem \eqref{eq:hard-input-lasso}, for which the $(c, S, \calD\ominus\calD)$-NSP  becomes
  \begin{align}
    \norm{AX_S}_{2,col} < c \norm{AX_{S^c}}_{2,col} \label{eq:nsp-no-input}
  \end{align}
  for all nonzero matrices $A \in\bR^{n\times n}$ such that $-AX \in \calD\ominus\calD$. The NSP property with $c=1$ ensures that $\bar A$ is the unique solution to the optimization problem
  \begin{subequations}
  \begin{align}
    \min_{\substack{A \in \mathbb{R}^{n \times n}, \\ D\in \calD} }\;\, & \sum_{i=0}^{T-1} \norm{d_i}_2                           \label{eq:hard-lasso} \\
    s.t. \quad                & x_{i+1} = A x_i + d_i, \quad {i=0, \ldots, T-1},
  \end{align}    
  \end{subequations}
where the states $x_i, i\in \{0, \ldots, T\}$, are generated according to
  \begin{align*}
    x_{i+1} = \bar A x_i + \bar d_i, \quad {i=0, \ldots, T-1}.
  \end{align*}
  \vspace{-0.3em}
\end{remark}
\vspace{-0.8cm}
\section{THE NOISY CASE}
\label{sec:noisy}

This section studies the noisy case, where some of the disturbances $d_0, \dots, d_{T-1}$ represent regular noise values and others are engineered by an attacker. Let $S$ denote the attack times, meaning that $d_i$ represents an attack if $i \in S$. For $i \in S^c$, the parameter $d_i$ represents noise and its value is often small in practice. The next theorem provides an error bound for estimating the matrices $A$ and $B$.

\begin{theorem}~\label{thm:error}
  Assume that  $T>(m+n)$ and that the matrix $\begin{bmatrix}
      X \\ U
    \end{bmatrix}$ has full row rank. If $(X,U)$ satisfies the $(c, S, \calD\ominus\calD)$-NSP with $c<1$, then each solution $(\hat A, \hat B, \hat D)$ to the optimization problem~\eqref{eq:hard-input-lasso} satisfies
  \begin{align*}
     & \norm{[\hat A - \bar A, \hat B-\bar B]}_F
    \leq 2\frac{1+c}{1-c} \times \frac{\norm{\bar D_{S^c}}_{2, col}}{\sigma_{\min}\left(    \begin{bmatrix}
          X \\ U
        \end{bmatrix} \right)}.
  \end{align*}
\end{theorem}
  The term $2\norm{\bar D_{S^c}}_{2, col}$ on the right-hand side of Theorem~\ref{thm:error} can be improved to $2\norm{\bar D_{S^c}}_{2, col} - \norm{\bar D}_{2, col} + \norm{\hat D}_{2, col}$ using similar techniques as those in basis pursuit; see for example \cite[Theorem 4.14]{foucartMathematicalIntroductionCompressive2013} that has proven an equivalence to $c$-NSP for basis pursuit problems. This term shows that the intensity of noise is zero in the noiseless case subject to attacks. Problem \eqref{eq:hard-input-lasso} is a special case of basis pursuit where measurements are correlated. The bound, including the constant $\frac{1+c}{1-c}$, could potentially be improved with more knowledge about the constraints (see \cite{eldarCompressedSensingTheory2012} for a similar scenario).
\section{SATISFACTION OF NSP}
\label{sec:nsp}

After observing the states and input sequence, condition \eqref{eq:csnsp} enables certifying whether one can recover the true dynamics using problem \eqref{eq:hard-input-lasso}. Theorem~\ref{thm:error} has shown that the NSP condition is useful in obtaining a bound o the identification error. The following lemmas attempt to derive stronger conditions that are more tractable than $(c, S, \calD)$-NSP. They can be combined with the results of the previous two sections to understand how to design the input to improve the likelihood of successfully recovering the system matrices through the convex optimization problem \eqref{eq:hard-input-lasso}.
\begin{lemma}\label{lem:singular-value-nsp}
  If $T\geq(m+n)$ and \begin{align}
    \sqrt{|S|}\sigma_{\max}\begin{bmatrix}
      X_S \\U_S
    \end{bmatrix} < c \times \sigma_{\min}\begin{bmatrix}
      X_{S^c} \\ U_{S^c}
    \end{bmatrix} \label{eq:sigma-nsp},
  \end{align}
  where $S=\supp(\bar D)$ and $\abs{S^c} \geq m+n$, then $\begin{bmatrix} X^T &  U^T\end{bmatrix}^T$ satisfies the $(c, S, \mathcal R)$-NSP for every range set $\mathcal{R}$.
\end{lemma}
\begin{definition}
  Given a matrix $V = [v_0, \ldots, v_{T-1}]$ and a natural number $s$, $V$ is said to be $s$-self-decomposable if for all indices $I \subseteq \{0, 1,  \ldots, T-1\}$ of size $\abs{I}=s$, it holds that $ V_{i}\in \range(V_{\notin I})$ for all $i\in I.$ The $s$-self-decomposable amplitude is defined as
  \begin{align}
    \xi_s(V) := \max_{\stackrel{I\subseteq \{0, \ldots, T-1\}}{\abs{I}=s}} \min_{\stackrel{\Gamma_I \in \bR^{(T-s)\times s}}{\Gamma_I = [\gamma_i]_{i\in I}}} \left\{\sum_{k\in I}\norm{\gamma_k}_{\infty}: V_{I} = V_{\notin I} \Gamma_I \right\}. \label{eq:decomposable-def}
  \end{align}
\end{definition}
If $U$ is $s$-self-decomposable, by definition it is also $t$-self-decomposable for $t<s$. We are particularly interested in the cases when $s=1$ and $s=\abs{S}$. 

\begin{lemma} \label{lem:xi-s-decomposable}
  If $\begin{bmatrix}
      X^T & U^T
    \end{bmatrix}^T$ has full row rank and is $s$-self-decomposable where $s=\abs{S}$, then it satisfies the $(c, S, \mathcal R)$-NSP for every $c>\xi_s(\begin{bmatrix}
        X^T & U^T
      \end{bmatrix}^T)$.
\end{lemma}

\begin{lemma} \label{lem:xi-1-decomposable}
  Given $S=\supp(\bar D)$ with $\abs{S}>1$, assume that $\begin{bmatrix}
      X^T & U^T
    \end{bmatrix}^T$ has full row rank and is $1$-self-decomposable where \begin{align*}
    \xi_1 := \xi_1(\begin{bmatrix}
        X^T & U^T
      \end{bmatrix}^T)  \leq \frac{1}{\abs{S}-1}.
  \end{align*}
  Then, it satisfies the $(c, S, \mathcal R)$-NSP and \[c > \frac{|S|\xi_1}{1-(\abs{S}-1)\xi_1}.\]
\end{lemma}

\begin{remark}
  Lemma~\ref{lem:xi-1-decomposable} implies that when $\xi_1 < \frac{1}{2\abs{S}-1}$, the ground truth $(\bar A, \bar B, \bar D)$ is recoverable through problem \eqref{eq:hard-input-lasso}.
\end{remark}

We have yet not answered how the control input sequence $u_0, \dots, u_{T-1}$ affects the satisfaction of NSP. This will be achieved after we consider a probabilistic model for the disturbances.

\section{A PROBABILISTIC MODEL}
\label{sec:prob}
The results of the previous section are applicable only when the state and input sequences have been observed --- they cannot be directly used to find a concrete input design scheme that achieves exact recovery in the noiseless case or asymptotic recovery in the noisy case. This section considers a particular type of random input.
Our approach relies on the observation that, despite the attacker's attempt, one can apply the block martingale small ball condition \cite{simchowitzLearningMixingSharp2018a} to obtain a probabilistic estimate on $\sigma_{\min}\left(\begin{bmatrix}
      X^T & U^T
    \end{bmatrix}^T\right)$.

We first restate the block martingale small ball (BMSB) condition \cite{simchowitzLearningMixingSharp2018a}. Define the filtration $\calF_t$ as 
the smallest $\sigma$-algebra obtained by the data available up to time $t$, i.e. $x_0, \ldots, x_t, u_0, \ldots, u_{t}, d_0, \ldots, d_{t-1}$, so that the vector-valued process 
$\begin{bmatrix}
    x_t^T & u_t^T
  \end{bmatrix}^T$ with $ t\geq 0$ is $\{\calF_t\}_{t\geq 0}$ adapted.
\begin{definition}
  Given a filtration $\{\calF_t\}_{t\geq 0}$ and a vector-valued process $V_t \in \bR^{d}$ with $ t\geq 0$, the process is said to satisfy $\left(k, \Gamma_{sb}, p\right)$-BMSB for some matrix $\Gamma_{sb}\succ 0$ if \begin{align*}
    \frac{1}{k}\sum_{i=1}^{k} \bP\left(\abs{\langle w, V_{j+i}\rangle}^2 \geq w^\top \Gamma_{sb}w | \calF_j \right) \geq p, \text{ almost surely}
  \end{align*}
  for all fixed vectors $w\in \bR^{d}$ with $\norm{w}_{2}=1$ and all $j\geq0$.
\end{definition}
We make the following two assumptions about the input and attack model. They ensure that the process is Gaussian.
\begin{assumption} \label{ass:gaussian-input}
  The input sequence $u_0, \dots, u_{T-1}$ are independent and identically distributed Gaussian random vectors with $N(0, \sigma^2 \mathbf{I}_m)$. 
\end{assumption}

\begin{assumption}\label{ass:attack-model}
  The attack model satisfies the following:
  \begin{itemize}
    \item The set of attack times $S\subseteq \{0, \ldots, T-1\}$ is fixed.
    \item For every $t\notin S$, $d_t$ is some noise that follows the distribution $N(0, \epsilon^2 \mathbf{I}_n)$ for some positive number $\epsilon$.
    \item For every $t\in S$, $d_t$ can be expressed as $ P x_t + Q u_t + e_t$, where $P$ and $Q $ are constant matrices of compatible size. 
    The matrices $P$ and $Q$ are not dependent on $\calF_t$. Also, $e_t$ is a random variable that follows the Gaussian distribution $N(0, \epsilon^2 \mathbf{I}_n)$ and is independent of $\calF_t$.
  \end{itemize}
\end{assumption}

The attack dynamics cannot be augmented with the system's dynamics since $d_t = 0$ if $t \not \in S$. It is desirable to bound the Gramian matrix, which is identified as a key measure of sample complexity in \cite{simchowitzLearningMixingSharp2018a}. Define the following constants:
\begin{align*}
  \alpha_{\min} & =\min(\sigma_{\min}(A+P),\sigma_{\min}(A))     \\
  \alpha_{\max} & =\max(\sigma_{\max}(A+P),\sigma_{\max}(A))     \\
  \beta_{\max}  & = \max(\sigma_{\max}({B+Q}),\sigma_{\max}(B)).
\end{align*}
\begin{lemma} \label{lem:lower-gramian}
  Let $ \Gamma_t = \bE \left[x_t x_t^\top\right] $ for $t = 0, \dots, T-1$. We have
  \begin{align*}
    \Gamma_t & \succeq \alpha_{\min}^2 \Gamma_{t-1} + \epsilon^2 \mathbf{I}_n,                                \\
    \Gamma_t & \preceq \alpha_{\max}^2 \Gamma_{t-1} + \left(\epsilon^2 + \beta_{\max}^2\right) \mathbf{I}_n.
  \end{align*}
  In particular, 
  \begin{align*}
    \Gamma_t & \succeq \sum_{i=0}^{T-1} \alpha_{\min}^{2i} \epsilon^2 \mathbf{I}_n   \\
    \Gamma_t & \preceq \Gamma_{t}^{\max} := \alpha_{\max}^{2t} \Gamma_0 + \sum_{i=0}^{T-1} \alpha_{\max}^{2i} \left(\epsilon^2 + \beta_{\max}^2\right) \mathbf{I}_n.
  \end{align*}
\end{lemma}

Let $\Gamma := \text{diag}(\epsilon^2 \mathbf{I}_n, \sigma^2 \bf I_m)$. The next lemma confirms that the BMSB condition can be leveraged for our problems.
\begin{lemma}\label{lem:random-bmsb} Under Assumptions \ref{ass:gaussian-input} and \ref{ass:attack-model}, for every sequence of indices $0\leq s_0 < s_1 < s_2, \ldots$,
  the sub-process $    \begin{bmatrix}
      x_{s_t}^T & u_{s_t}^T
    \end{bmatrix}^T$ with $ t\geq0$ satisfies the $(k, \frac12\Gamma, \frac1{12})$-BMSB condition.
\end{lemma}
The BMSB condition provides a non-asymptotic bound on the singular value of $\begin{bmatrix}
    X^T & U^T
  \end{bmatrix}^T$.

\begin{proposition}\label{prop:bound-sigma} Under Assumptions \ref{ass:gaussian-input} and \ref{ass:attack-model}, define $C(I) := \left(m \sigma^{2}\abs{I}  + \sum_{i\in I} \\ 
 \text{tr}(\Gamma^{\max}_{i})\right)$, where $\Gamma^{\max}_{i}$ is given in Lemma~\ref{lem:lower-gramian}. For every subset $I\subseteq \{0, 1, \ldots, T-1\}$, we have
  \begin{align}
    \bP\left(\sigma_{\max}\left(\begin{bmatrix}
        X_{I} \\ U_{I}
      \end{bmatrix}\right)> \sqrt{\frac{C(I)}{\eta}} \right) & \leq \eta \label{eq:maxbound}
  \end{align}
  and
  \begin{align}
     & \bP\left(\sigma_{\min}\left(\begin{bmatrix}
        X_{I} \\ U_{I}
      \end{bmatrix}\right)<  \min(\epsilon, \sigma)\sqrt{\frac{k\lfloor \abs{I}/k\rfloor p^2}{16}}\right)                  \leq \eta \nonumber \\
     & + \exp\left(-\frac{\abs{I}p^2}{10k}  + 2(m+n) \log(10/p)  \right.                                               \nonumber                                                         \\
     & \left.+ \frac12(m+n)\log\left(\frac{C(I)}{\min(\epsilon, \sigma)^2\frac{k\lfloor \abs{I}/k\rfloor p^2}{16} \eta^2}\right)\right)\label{eq:minbound}
  \end{align}
\end{proposition}
The proof is a direct consequence of the covering argument in \cite[Section D]{simchowitzLearningMixingSharp2018a}. We are now able to provide a sufficient condition for the satisfaction of NSP in our attack model.
\begin{theorem}\label{thm:nsp-gaussian-satisfy}
  Assume that $\alpha_{\max}<1$. Given  $c, \eta>0$, there exist constants $N$ and $h>0$ such that $\begin{bmatrix}
      X^T & U^T    \end{bmatrix}^T$ is c-NSP with probability at least $1-3\eta$ as long as $|S|^2 < h |S^c|$ and $\abs{S^c} > N$.
\end{theorem}

\section{NUMERICAL EXPERIMENTS}

\begin{figure}
    \centering
    \includegraphics[scale =0.15]{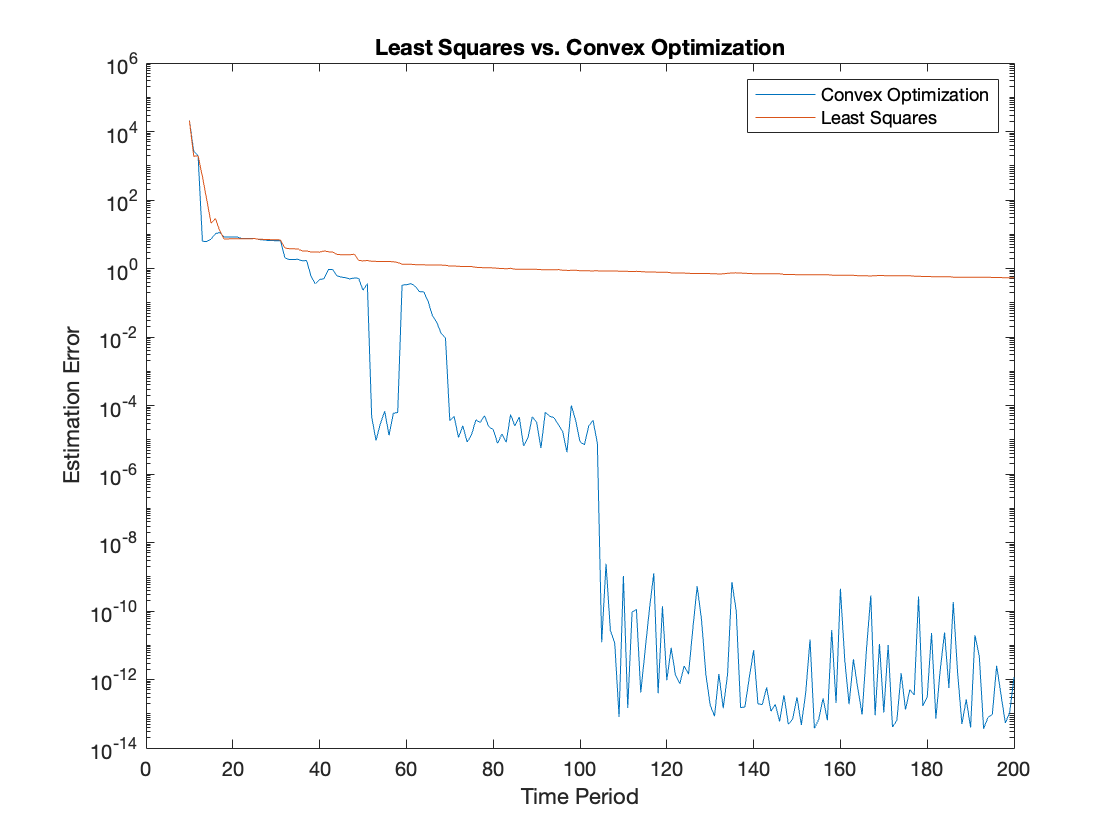}
    \caption{Estimation Error Comparison for Noiseless Case with $\bar A \in \mathbb{R}^{10 \times 10}$ with Time Horizon $T = 200$}
    \label{fig:exp1}
\end{figure}

\begin{figure}
    \centering
    \includegraphics[scale =0.15]{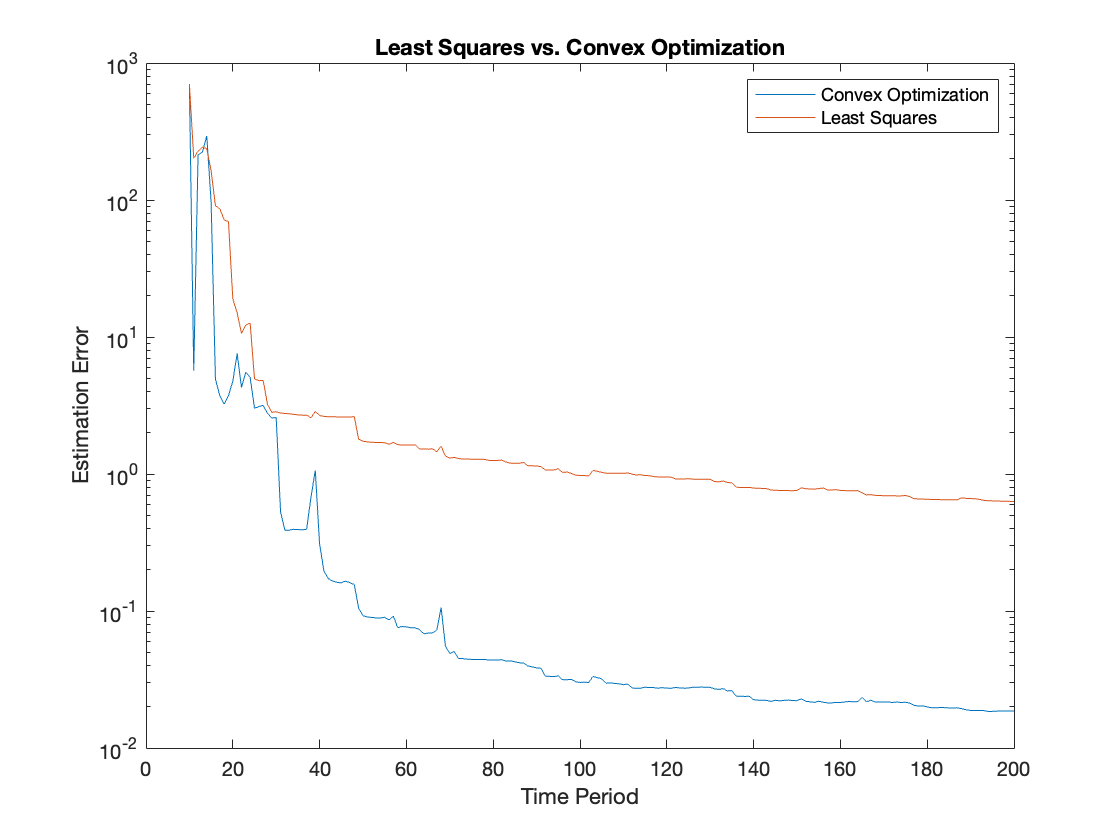}
    \caption{Estimation Error Comparison for Noisy Case with $\bar A \in \mathbb{R}^{10 \times 10}$ with Time Horizon $T = 200$}
    \label{fig:exp2}
\end{figure}

We provide numerical experiments to support the theoretical results obtained in this paper. For illustration purposes, we focus on an autonomous system of the form $x_{t+1} = \bar A x_t + \bar d_t$. We generate a diagonal matrix $\Lambda$ whose diagonal entries are uniformly distributed between 0 and 1. In addition, we generate a Gaussian matrix $P$ whose entries are independent and identically distributed  (i.i.d.) with $N(0,1)$. Then, we set the matrix $\bar A $ to be $ P \Lambda P^{-1}$. The vector $x_0$ is generated randomly based on a Gaussian distribution with i.i.d $N(0,1)$ entries. At each time instance, an adversarial attack occurs with the probability $p = 0.3$. An adversarial attack $d_i$ at time $i$ is distributed as $N(0, 100 \mathbf{I}_n)$. In the noisy case, a Gaussian noise vector with i.i.d. $N(0,1)$ entries are added at each time instance while they are omitted in the noiseless case.  We compare the estimation error of problem \eqref{eq:hard-input-lasso} with the estimation error of the least-squares problem:
\begin{align*}
  \min_{A \in \mathbb{R}^{n \times n}} \sum_{t=0}^{T-1}\norm{x_{i+1} - A x_i}_2^2.
\end{align*}
Given an estimate $\hat A$, the error is calculated as $\| \hat A - \bar A\|_F$. We plot the estimation errors of problem \eqref{eq:hard-input-lasso} and the least-squares method with respect to time. In Figure \ref{fig:exp1}, we report the results for the noiseless case. It is observed that after a sufficient number of data points are obtained, problem \eqref{eq:hard-input-lasso} exactly identifies the system. On the other hand, the least-squares estimation errors reach a plateau due to adversarial attacks.
In Figure \ref{fig:exp2}, we implement a similar analysis for the noisy case. It is seen that the estimation errors decrease significantly faster than those for the least-squares estimation when our convex optimization formulation is used. Thus, we can conclude that problem \eqref{eq:hard-input-lasso} provides a more accurate estimation of the system dynamics.

\section{CONCLUSION}
This paper studies an $l_1$-based identification scheme for a fully observable LTI system affected by sparse state disturbances. We find that as long as the attack is not too frequent, even assuming that the attack can take the form of a linear state and input feedback, an accurate state-space representation can be obtained. We derive some inequalities in the form of the null space property serving as conditions for the exact recovery of the model and develop a bound on the estimation error. It is intriguing to study when consistency and error bounds hold for other models of attack. More generally, other identification schemes such as iterative re-weighted least-squares and their variations are promising to analyze in the system identification context. 

\bibliographystyle{IEEEtran}
\bibliography{root}

\appendix

\section{APPENDICES}

\subsection{Proof of Theorem \ref{thm:nsp-unique}}

\begin{proof}
  Let ($A$, $B$, $D$) be any feasible solution to problem \eqref{eq:hard-input-lasso}. We will show that if the matrices are not equal to the ground truth $(\bar A, \bar B, \bar D)$, then they cannot be an optimal solution. The feasibility can be written as: 
  \begin{align}
    x_{i+1} & = \bar A x_i + \bar B u_i + \bar d_i, \quad {i=0, \ldots, T-1}, \nonumber \\
    x_{i+1} & = A x_i + B u_i + d_i, \quad {i=0, \ldots, T-1}. \nonumber
  \end{align}
  Taking the difference of the two equalities, $(\bar A - A, \bar B- B, \bar D - D)$ lies in the null space in the sense that
  \begin{align*}
    0 = (\bar A - A) x_i + (\bar B - B) u_i + (\bar d_i - d_i), \quad {i=0, \ldots, T-1},
  \end{align*}
  which can be written in matrix form as
  \begin{subequations}
      \begin{align}
    0 & = (\bar A - A) X_S + (\bar B - B) U_S + (\bar D_S - D_S) \label{eq:10a} \\
    0 & = (\bar A - A) X_{S^c} + (\bar B - B) U_{S^c} - D_{S^c}, \label{eq:10b}
  \end{align}
  \end{subequations}
  where $S = \supp(\bar D)$. Note that $\bar D - D \in \calD\ominus\calD$. If $A-\bar A=0$ and $B-\bar B=0$, then $D=\bar D$. If $D \neq \bar D$, then $A-\bar A$ and $B-\bar B$ are not both zero simultaneously. We apply the null space property to obtain
  \begin{align*}
     & \norm{D}_{2, col}  =  \norm{D_S}_{2, col} + \norm{-D_{S^c}}_{2, col}                                 \\
     & = \norm{D_S}_{2, col} + \norm{(\bar A - A) X_{S^c} + (\bar B - B) U_{S^c}}_{2, col}                  \\
     & > \norm{D_S}_{2, col} + \norm{(\bar A - A) X_{S} + (\bar B - B) U_{S}}_{2, col} \quad \text{(1-NSP)} \\
     & =  \norm{D_S}_{2, col} + \norm{(\bar D-D)_{S}}_{2, col}                                              \\
     & \geq \norm{\bar D_S}_{2, col}\quad \text{(triangle inequality)}                                      \\
     & = \norm{\bar D}_{2, col} \text{ (sparsity of disturbance)}.
  \end{align*}
  This means that $(A, B, D)$ is not an optimal solution to problem \eqref{eq:hard-input-lasso}. 

We also intend to show that the converse is true. Consider problem \eqref{eq:hard-input-lasso} with the state-space matrices $(A^*, B^*) = (\pi_S(A-\bar A), \pi_S(B-\bar B))$. Since we assume $(ii)$ is true, we know that $(\pi_S(A-\bar A), \pi_S(B-\bar B)), \pi_S(D-\bar D)) $ is the unique solution. Note that by \eqref{eq:10a} and \eqref{eq:10b}, the negative of $(\pi_{S^c}(A-\bar A), \pi_{S^c}(B-\bar B)), \pi_{S^c}(D-\bar D)) $ is also a feasible solution. By the optimality condition, we obtain that
   \begin{align*}
       \| \pi_S(D-\bar D)) \|_{2,col} & < \|  - \pi_{S^c}(D-\bar D)) \|_{2,col}, \\
       \| D_S-\bar D_S\|_{2,col} & < \| D_{S^c}-\bar D_{S^c} \|_{2,col}, \\
       \left \| [A', B']    \begin{bmatrix}
        X_{S} \\ U_{S}
      \end{bmatrix}   \right\|_{2, col}  & < \left \|[A', B']     \begin{bmatrix}
        X_{S^c} \\ U_{S^c}
      \end{bmatrix}   \right\|_{2, col},
   \end{align*}
   where $A' = A - \bar A$ and $B' = B - \bar B$. Since $A' \not = 0 $ and $B' \not = 0$ whenever $A$ and $B$ are different from $\bar A$ and $\bar B$, respectively, $(1, S, \calD\ominus\calD)$-NSP property holds.
\end{proof}

\subsection{Proof of Theorem \ref{thm:error}}

\begin{proof}
  The optimality of the solution implies that $\norm{\hat D}_{2, col} \leq \norm{\bar D}_{2, col}$. The constraints imply that
  \begin{align*}
    0 = (\bar A - \hat A)X + (\bar B - \hat B) U + (\bar D- \hat D).
  \end{align*}
  Also, $c$-NSP yields that

  \begin{equation}
    \norm{\bar D- \hat D}_{2, col} < (1+c) \norm{\bar D_{S^c}- \hat D_{S^c}}_{2, col}. \label{eq:nsp2}
  \end{equation}
  One can write:
  \begin{align*}
     & \norm{\bar D}_{2, col}  \geq \norm{\hat D}_{2, col}                       \\
     & =\norm{\bar D_S + (\hat D_S - \bar D_S)}_{2, col}  +  \norm{\bar D_{S^c} + (\hat D_{S^c} - \bar D_{S^c})}_{2, col}                                \\
     & \geq \norm{\bar D_S}_{2, col} - \norm{\hat D_S - \bar D_S}_{2, col} -  \norm{\bar D_{S^c}}_{2, col} + \norm{\hat D_{S^c} - \bar D_{S^c}}_{2, col} \\
     & \geq \norm{\bar D_S}_{2, col} - \norm{\bar D_{S^c}}_{2, col} + (1-c)\norm{\hat D_{S^c} - \bar D_{S^c}}_{2, col}                                   \\
     & \geq \norm{\bar D_S}_{2, col} - \norm{\bar D_{S^c}}_{2, col} + \frac{1-c}{1+c}\norm{\hat D - \bar D}_{2, col}
  \end{align*}
  where we have used the triangle inequality together with NSP inequalities. Cancelling $\norm{\bar D_{S}}_{2, col}$ on both sides, we obtain
  \begin{align*}
    \norm{\hat D - \bar D}_{2, col} \leq 2\frac{1+c}{1-c}\norm{\bar D_{S^c}}_{2, col}.
  \end{align*}
  The bound above can be translated to a bound on $(A, B)$ through the matrix norm inequality (note that $T\geq (m+n)$):
  \begin{align*}
     & \norm{[\hat A - \bar A, \hat B-\bar B]}_F\sigma_{\min}\left(    \begin{bmatrix}
        X \\ U
      \end{bmatrix} \right)                      \\
     & \leq \norm{\hat D - \bar D}_{F} \leq \norm{\hat D - \bar D}_{2, col} \leq 2\frac{1+c}{1-c}\norm{\bar D_{S^c}}_{2, col}.
  \end{align*}
\end{proof}

\subsection{Proof of Lemma \ref{lem:singular-value-nsp}}

\begin{proof}
  For two matrices $A, B$ such that $-AX - BU \in \mathcal R$ and $(A,B)$ are not both zero, we can upper-bound and lower-bound the norms: 
  \begin{align*}
    \left \|[A, B]    \begin{bmatrix}
        X_{S} \\ U_{S}
      \end{bmatrix}   \right\|_{2, col}  & \leq  \sqrt{|S|} \left \|[A, B]    \begin{bmatrix}
        X_{S} \\ U_{S}
      \end{bmatrix}   \right\|_F                         \\
  & \leq  \sqrt{|S|} \norm{[A, B]}_{F}  \sigma_{\max}\left(\begin{bmatrix}
        X_{S}^T & U_{S}^T
      \end{bmatrix}^T \right) \\
    \left\|[A, B]     \begin{bmatrix}
        X_{S^c} \\ U_{S^c}
      \end{bmatrix}  \right\|_{2, col} & \geq \left \|[A, B]     \begin{bmatrix}
        X_{S^c} \\ U_{S^c}
      \end{bmatrix}  \right\|_{F}                                  \\
    & \geq  \norm{[A, B]}_F  \sigma_{\min}\left(\begin{bmatrix}
        X_{S^c}^T & U_{S^c}^T
      \end{bmatrix}^T  \right),
  \end{align*}
  where we use the relationship between $(2,col)$-norm and Frobenius norm. The last inequality uses the assumption that $\abs{S^c} \geq m+n$.
  The inequality \eqref{eq:sigma-nsp} therefore implies $1$-NSP. The last statement follows from Theorem~\ref{thm:nsp-unique} by setting $\mathcal R = \calD\ominus\calD$ and $c=1$.
\end{proof}

\subsection{Proof of Lemma \ref{lem:xi-s-decomposable}}

\begin{proof}
  Assuming that $\begin{bmatrix}
      X \\ U
    \end{bmatrix}$ is $s$-self-decomposable where $s=\abs{S}$, one can find a matrix $\Gamma^*_S = [\gamma^*_i]_{i\in S}$ that is the minimizer of the inner optimization problem \eqref{eq:decomposable-def}:
  \begin{align*}
     & \left\|[A,B] \begin{bmatrix}
        X_S \\ U_S
      \end{bmatrix}\right\|_{2,col} =  \left\|[A,B] \begin{bmatrix}
        X_{S^c} \\ U_{S^c}
      \end{bmatrix} \Gamma_S^*\right\|_{2,col} \\
     & \leq \sum_{i\in S} \norm{\gamma_i^*}_{\infty}\left\|[A,B] \begin{bmatrix}
        X_{S^c} \\ U_{S^c}
      \end{bmatrix}\right\|_{2,col}                  \\
     & \leq \xi_s \left(\begin{bmatrix}
        X \\ U
      \end{bmatrix}\right)\left\|[A,B] \begin{bmatrix}
        X_{S^c} \\ U_{S^c}
      \end{bmatrix}\right\|_{2,col}.
  \end{align*}
  It remains to obtain a strict inequality when we relax $c>\xi_s(\begin{bmatrix}
        X \\ U
      \end{bmatrix})$. To prove by contradiction, suppose that this does not yield a strict inequality. Then
      \begin{align*}
    [A,B] \begin{bmatrix}
      X_{S^c} \\ U_{S^c}
    \end{bmatrix} = 0 \text{ and } [A,B] \begin{bmatrix}
      X_{S} \\ U_{S}
    \end{bmatrix} = 0.
  \end{align*}
  Since $\begin{bmatrix}
      X^T & U^T
    \end{bmatrix}^T$ has full row rank, this means that $[A, B] = 0$.
\end{proof}

\subsection{Proof of Lemma \ref{lem:xi-1-decomposable}}

\begin{proof}
  For every $s\in S$, one can find a vector $\gamma_s^*$ that is the minimizer of the inner optimization Problem \eqref{eq:decomposable-def}:
  \begin{align*}
    \left\|[A,B] \begin{bmatrix}
        X_s \\ U_s
      \end{bmatrix}\right\|_{2} & = \left\|[A,B] \begin{bmatrix}
        x_s \\ u_s
      \end{bmatrix}\right\|_2                                   \\
  & = \left\|[A,B] \begin{bmatrix}
        X_{\neq s} \\ U_{\neq s}
      \end{bmatrix} \gamma_s^*\right\|_2                        \\
  & \leq  \norm{\gamma_s}_{\infty}\left\|[A,B] \begin{bmatrix}
        X_{\neq s} \\ U_{\neq s}
      \end{bmatrix}\right\|_{2,col} \\
   & \leq  \xi_1 \left\|[A,B] \begin{bmatrix}
        X_{\neq s} \\ U_{\neq s}
      \end{bmatrix}\right]\|_{2,col}
  \end{align*}
  Hence,
  \begin{align*}
    \left\|[A,B] \begin{bmatrix}
        X_s \\ U_s
      \end{bmatrix}\right\|_{2} \leq \frac{ \xi_1 }{1+ \xi_1 } \left\|[A,B] \begin{bmatrix}
        X \\ U
      \end{bmatrix}\right\|_{2,col}.
  \end{align*}
  Summing over $s\in S$, we obtain
  \begin{align*}
    \left\|[A,B] \begin{bmatrix}
        X_S \\ U_S
      \end{bmatrix}\right\|_{2, col} \leq |S|\frac{ \xi_1 }{1+ \xi_1 } \left\|[A,B] \begin{bmatrix}
        X \\ U
      \end{bmatrix}\right\|_{2,col}.
  \end{align*}
  After rearranging the terms, the proof is completed by noting that, as in the proof of Lemma~\ref{lem:xi-s-decomposable}, the full rank assumption implies that when we select $c> \frac{|S|\xi_1}{1-(\abs{S}-1)\xi_1}$ the inequality is strict for $(A, B)\neq 0$.
\end{proof}

\subsection{Proof of Lemma \ref{lem:lower-gramian}}

\begin{proof}
  By conditioning on $\calF_{t-1}$, we have

  \begin{equation}
    \begin{aligned}
                  \bE[x_{t}x_{t}^\top | \calF_{t-1}]
 = & \bE[(A x_{t-1} + B u_{t-1} + d_{t-1}) \\ \notag  & \times (A x_{t-1} + B u_{t-1} + d_{t-1})^\top | \calF_{t-1}].
    \end{aligned}\label{eq:theterm}
  \end{equation}

  We analyze two cases. When $t-1\in S$, the term \eqref{eq:theterm} becomes
          \begin{align*}
             & \bE[((A + P)x_{t-1} + (B+Q) u_{t-1} + e_{t-1}) \\ & \quad \times ((A + P)x_{t-1} + (B+Q) u_{t-1} + e_{t-1})^\top | \calF_{t-1}].
          \end{align*}
          Taking the expectation on both sides, we obtain
          \begin{align*}
             & \Gamma_t = \bE \left[x_t x_t^\top\right]                                                  \\
             & \stackrel{(b)}{=} (A+P) \Gamma_{t-1}(A+P)^\top + \sigma^2(B+Q)(B+Q)^\top  + \epsilon^2 I.
          \end{align*}
          where (b) follows by noting that $u_{t-1}$ and $e_{t-1}$ are independent of $x_{t-1}$ and have mean zero.
   When $t-1\notin S$, the term \eqref{eq:theterm} becomes \begin{align*}
             & \bE[(A x_{t-1} + B u_{t-1} + e_{t-1})                             \\
             & \quad \times (A x_{t-1} + B u_{t-1} + e_{t-1})^\top | \calF_{t-1}]. 
          \end{align*}
          Taking the expectation in a similar way yields that
          \begin{align*}
             & \Gamma_t = A\Gamma_{t-1}A^\top + \sigma^2 BB^\top + \epsilon^2 I.
          \end{align*}

  In both cases, we can lower-bound $\Gamma_t$ by leaving out the positive semi-definite term and using the minimal singular values of the multipliers:
  \begin{align*}
    \Gamma_t \succeq \alpha_{\min}^2 \Gamma_{t-1} + \epsilon^2 I.
  \end{align*}
  The upper bound follows similarly by bounding via the maximum singular values. The proof is completed by induction.
\end{proof}

\subsection{Proof of Lemma \ref{lem:random-bmsb}}

\begin{proof}
  For clarity of notation, we will prove the result for $s_t=t$.

  We will prove that the process is $3$-Paley-Zygmund~\cite[Lemma 3.9]{Simchowitz:EECS-2021-33} and conclude BMSB as a consequence, following a similar argument as in \cite{Simchowitz:EECS-2021-33}. Fix a vector $\begin{bmatrix}
      w \\ v
    \end{bmatrix} \in \bR^{n+m}$. Given fixed times $j\geq 0$ and $i\geq 1$, one can write:
  \begin{align*}
    x_{i+j} | \mathcal{F}_j = A^{i}x_j + \sum_{0\leq k\leq i-1} A^{i-k-1}(B u_{j+k} + d_{j+k})| \mathcal{F}_j
  \end{align*}
  We may substitute the expression of $d_{s+i}$ where $ 0\leq i \leq t-1$, and find that the conditional distribution of $\langle w, x_{t+s} \rangle + \langle v, u_{t+s}\rangle | \calF_s$ is Gaussian.
  Let $Y_i = (\langle w, x_{j+i} \rangle + \langle v, u_{j+i}\rangle)^2$ and $Z_{j+i} = \begin{bmatrix}
      x_{j+i} \\ u_{j+i}
    \end{bmatrix}$ for $i\geq 1$. It can be concluded that
  \begin{align*}
    \bE[Y_i | \calF_j] & = [w, v]^\top \bE[Z_{j+i}Z_{j+i}^\top | \calF_j]     \begin{bmatrix}
      w \\v
    \end{bmatrix}     \\
                       & \stackrel{(a)}{=} w^\top \bE[x_{j+i} x_{j+i}^\top | \calF_j] w  + \sigma^2 v^\top v \\
                       & \stackrel{(b)}{\geq}  \epsilon^2 w^\top w+ \sigma^2 v^\top v
  \end{align*}
  where (a) holds because $u_{j+i}$ is independent of $x_{j+i}$ and is Gaussian with the variance $\sigma^2$; (b) follows from the lower bound of Gramian in Lemma~\ref{lem:lower-gramian}. To evaluate the condition in BMSB, note that
  \begin{align}
     & \bP\left({\frac1k \sum_{i=1}^k Y_i \geq \frac12 \epsilon^2 w^\top w  + \sigma^2 v^\top v } \middle| \calF_j \right)      \nonumber                       \\
     & \geq \bP\left({\frac1k \sum_{i=1}^k Y_i \geq \frac12 \bE\left[\frac1k \sum_{i=1}^k Y_i | \calF_j\right]} \middle| \calF_j \right)   \nonumber                       \\
     & \stackrel{(c)}{\geq} \frac14 \frac{[\bE[\sum_{i=1}^{k} Y_i| \calF_j]]^2}{\bE\left[(\sum_{i=1}^{k} Y_i)^2 | \calF_j\right]} \label{eq:bmsb-bound-almost},
  \end{align}
  where (c) uses the Paley-Zygmund inequality. Since $Y_i|\calF_j$ takes the form of $Z^2$ where $Z$ is a Gaussian random variable, we have
  \begin{equation}
    \bE[Z^4] \leq 3 (\bE[Z^2])^2 \label{eq:3-paley},
  \end{equation}
  and therefore, 
  \begin{align*}
    \bE\left[\left(\sum_{i=1}^{k} Y_i \right)^2 | \calF_j\right] & = \sum_{i, i' =1}^k \bE\left[ Y_i Y_{i'}| \calF_j\right]                                  \\
 & \stackrel{(d)}{\leq} \sum_{i, i' =1}^k \sqrt{ \bE[Y_i ^2| \calF_j]\bE[Y_{i'}^2| \calF_j]} \\
  & \stackrel{(e)}{\leq}  3 \sum_{i, i' =1}^k\bE[Y_i| \calF_j] \bE[Y_{i'}| \calF_j],
  \end{align*}
  where
  (d) uses the Cauchy inequality and (e) follows from \eqref{eq:3-paley}. Combining this inequality with~\eqref{eq:bmsb-bound-almost} completes the proof of the BMSB condition.
\end{proof}

\subsection{Proof of Theorem \ref{thm:nsp-gaussian-satisfy}}

\begin{proof}
  When $\alpha_{\max}<1$, Lemma \ref{lem:lower-gramian} shows that $\text{tr}(\Gamma_{i}^{\max})$ can be bounded and hence $C(I) = O(\abs{I})$. Applying Proposition~\ref{prop:bound-sigma} for $I = S$ and $I=S^c$, respectively, one can conclude that there exist constants $N, c'$, and $c'''$ that do not depend on $S$, such that when $\abs{S^c}>N$, with probability as least $1-3\eta$, the following two conditions hold:
  \begin{align*}
    \sigma_{\max}\left(\begin{bmatrix}
        X_{S} \\ U_{S}
      \end{bmatrix}\right) & \leq c' \sqrt{\frac{\abs{S}}{\eta}}, \quad
    \sigma_{\min}\left(\begin{bmatrix}
        X_{S^c} \\ U_{S^c}
      \end{bmatrix}\right) & \geq c''\sqrt{\abs{S^c}}.
  \end{align*}
  Therefore, one can select a small enough $h>0$ such that when  $|S|^2 < h |S^c|$,  \begin{align}
    \sqrt{|S|}\sigma_{\max}\begin{bmatrix}
      X_S \\U_S
    \end{bmatrix} < c \cdot \sigma_{\min}\begin{bmatrix}
      X_{S^c} \\ U_{S^c}
    \end{bmatrix}
  \end{align}
   holds with probability at least $1-3\eta$. Lemma~\ref{lem:singular-value-nsp} then applies and we conclude that  $\begin{bmatrix}
      X \\U
    \end{bmatrix}$ is c-NSP.
\end{proof}

\end{document}